\documentclass[nofootinbib,prd,twocolumn,showpacs,showkeys,preprintnumbers]{revtex4-1}
\usepackage{hyperref,amssymb,amsmath,mathrsfs,bm,graphicx}
\usepackage{color}
\begin{document}
\title { ELECTROMAGNETIC  RADIATION PRODUCES FRAME DRAGGING }
\author{L. Herrera}  
\email{laherrera@cantv.net.ve}
\affiliation{Departamento de F\'\i sica Te\'orica e Historia de la Ciencia,
Universidad del Pa\'{\i}s Vasco, Bilbao, Spain}
\altaffiliation{Also at U.C.V., Caracas}

\author{W. Barreto} 
\affiliation{Centro de  F\'{\i}sica Fundamental, Facultad de Ciencias, Universidad de Los Andes, M\'erida, Venezuela}
\email{wbarreto@ula.ve}

\date{\today}
\begin{abstract}
It is shown that   for a generic electrovacuum spacetime,  electromagnetic radiation  produces vorticity of worldlines of observers in a Bondi--Sachs frame.  Such an effect (and the ensuing gyroscope precession with respect to the lattice)  which is a reminiscence of  generation of  vorticity  by gravitational radiation, may be  linked to the nonvanishing of  components of the Poynting  and the super--Poynting vectors on the planes othogonal to the vorticity vector. The possible observational relevance of such an effect is commented.
\end{abstract}
\date{\today}
\maketitle
\section{Introduction}

{Nowadays high--accuracy experiments in space can test relativistic gravity. These endeavours surely should lead to direct detection of gravitational waves \cite{Turyshev_etal}. Particularly interesting are the frame dragging observations using cryogenic gyroscopes in Gravity Probe B searches \cite{22} and from data provided by LAGEOS satellites \cite{ciu1, ciu2}. These successful experiments have fundamental implications for general relativity since frame dragging is one of the most conspicuous general relativistic effect \cite{shiff}. Furthermore, they are also  very important in astrophysics because the frame dragging has been invoked as a mechanism to drive relativistic jets emanating from galactic nuclei \cite{thorne}. Dual jets are expected for inspiral supermassive binary black hole systems which produce prodigious quantities of gravitational waves and energetic electromagnetic events \cite{pll10}, which in some way have to be closely related. Less impressive outcomes but of utmost importance are the  intense electromagnetic outbursts from hyperenergetic phenomena such as collapsing hypermassive neutron stars \cite{21} and Gamma Ray Bursts \cite{20}.}

Now, it is a well established fact that gravitational radiation produces vorticity in the congruence of observers with respect to the compass of inertia  \cite{1}--\cite{5}. Since the vorticity  vector decribes the proper angular velocity of the compass of inertia (gyroscope) with respect to reference particles \cite{6}, it is clear that a frame dragging effect is associated with gravitational radiation. 

In \cite{5} it was further shown that such a vorticity is closely related to the non--vanishing of super-Poynting vector components on the plane orthogonal to the vorticity vector.  Later it was shown that the vorticity appearing in stationary vacuum spacetimes is also related to the existence of a flow of superenergy on the plane orthogonal to the vorticity vector \cite{7}.

 The rationale to link  vorticity and the super-Poynting vector comes from an idea put forward by  Bonnor in order to explain the appearance of vorticity in the spacetime generated by a charged magnetic dipole \cite{8}. Indeed, Bonnor observes that for such a system there exists  a non--vanishing component of the Poynting vector, describing a flow of electromagnetic energy round in circles \cite{9}. He then suggests that such a circular flow of energy affects inertial frames by producing vorticity of congruences of particles, relative to the compass of inertia. Later, this conjecture was shown to be valid for a general axially symmetric stationary electrovacuum metric \cite{10}. 

In \cite{4} it was  suggested for the first time that a similar mechanism might be at the origin  of vorticity in the gravitational case, i.e.  a circular flow of gravitational energy would produce  vorticity. However due to the well known problems associated  to  a local and invariant  definition of  gravitational energy it was unclear at that time what expression for the ``gravitational'' Poynting vector  should be used. Following a suggestion by Roy Maartens to one of us (L.H.) we tried in \cite{5} with the super--Poynting vector based on the Bel--Robinson tensor \cite{11}--\cite{14}. Doing so  we were able to establish the link between  gravitational radiation and vorticity, invoking   a mechanism similar to that proposed by Bonnor for the charged magnetic dipole.

Our purpose with this work is  to tie up an important  loose end related to this problem, namely: Does electromagnetic radiation produces vorticity?, and if so can we explain such vorticity through a Bonnor--like  mechanism?

We shall analyze a generic electrovacuum spacetime following the scheme {developed} by van der Burg \cite{15}, which is an extension of the Bondi  formalism \cite{16, 17}  as to include electromagnetic fields. After a brief review of van der Burg paper we shall proceed to calculate the vorticity, the Poynting and the super--Poynting vectors. Based on the analysis of the obtained results we shall then {answer the two} questions raised above.

\section{Asymptotic expansions for the Einstein--Maxwell field}
In \cite{15} van der Burg presents a generalization of the Bondi--Sachs formalism for the  Einstein--Maxwell system.
This has, among other things, the virtue of providing a clear and precise criterion for the existence of gravitational and electromagnetic  radiation. Namely, if the news functions (gravitational and/or electromagnetic)  are zero over a time interval, then there is no radiation (gravitational and/or electromagnetic) during that interval. 

The formalism has as its main drawback \cite{18} the fact that it is based on a series expansion which could not give closed solutions and which raises unanswered questions about convergence and appropriateness of the expansion. However since we shall consider regions of spacetime   very far from the source, we shall use in our calculations only the leading terms in the expansion of metric functions. Furthermore, since the source is assumed to radiate during a finite interval, then no problem of convergence appears \cite{19}.

The general form of the metric in the Bondi--Sachs formalism can be written as \cite{17}
\begin{eqnarray}
ds^2&=&\left(e^{2\beta}\frac{V}{r}-r^2h_{AB}U^AU^B\right)du^2 
+2e^{2\beta}dudr\nonumber \\&+&2r^2h_{AB}U^Bdudx^A-r^2h_{AB}dx^Adx^B ,\label{metric}
\end{eqnarray}
where all the metric components are functions of  $x^0=u$, $x^1=r$, $x^2=\theta$, $x^3=\phi$. $\beta$ is the expansion of outward null rays, $V$ the Newtonian like potential. If we make a $2+1$ foliation doing constant $r$, $e^{2\beta}V/r$ is the lapse, $U^A$ the shift and $h_{AB}$ the 2--surface metric of constant $u$, satisfying $h^{AB}h_{BC}=\delta^A_C$. $u$ is a timelike  coordinate, constant along outgoing radial null geodesics, while $r$ is a {luminosity--distance} parameter. 

The electromagnetic field is characterized by a skew--symmetric tensor $F_{\mu\nu}$ giving rise the energy--momentum tensor $T_{\mu\nu}$  defined by
\begin{equation}
T_{\mu\nu}=\frac{1}{4}g_{\mu\nu}F_{\gamma\delta}F^{\gamma\delta}-g^{\gamma\delta}F_{\mu\gamma}F_{\nu\delta} ,
\end{equation}
being $g_{\mu\nu}$ the metric tensor given by (\ref{metric}). The Einstein--Maxwell field equations for {the electro--vacuum} spacetime outside the source are
\begin{eqnarray}
R_{\mu\gamma} + T_{\mu\gamma} = 0,\\
F_{[\mu\nu,\delta]}=0,\\
F^{\mu\nu}_{;\nu}=0,
\end{eqnarray}
where $R_{\mu\gamma}$ is the Ricci tensor. {Note that the units used are $16\pi G=c=1$.} Our purpose here is to extract general information from the radiative zone, at $\mathcal{J}^+$ ({future null--infinity}), where the well known asymptotic expansion in power of $r^{-1}$ of Bondi seems to be the most convenient and precise. 

The metric (\ref{metric}) can be written as follows \cite{vbn}
\begin{eqnarray}
ds^2&=&(Vr^{-1}e^{2\beta}-r^2 e^{2\gamma}U^2\cosh
2\delta -r^2 e^{-2\gamma} W^2\times\nonumber\\
 &&\cosh 2\delta
- 2r^2 UW\sinh 2\delta)du^2
 +2e^{2\beta}dudr
 +2r^2\times\nonumber\\
 &&(e^{2\gamma}U\cosh 2\delta
+W \sinh 2\delta)dud\theta
 +2r^2(e^{-2\gamma}W\times\nonumber\\
&&\cosh 2\delta + U\sinh
2\delta) \sin\theta dud\phi -r^2(e^{2\gamma}\cosh 2\delta d\theta^2\nonumber\\
&&+e^{-2\gamma}\cosh 2\delta \sin^2\theta d\phi^2
 +2\sinh 2\delta \sin\theta d\theta d\phi),\,\,\,\,\,\;\;\;\;\label{bsm}
\end{eqnarray}
if we choose the gauge of Bondi, $\det(h_{AB})=\det(q_{AB})$, where
$q_{AB}$ is the unit 2--sphere metric; $\gamma$, $\delta$, $U$, $W$ are functions of $(u,r,\theta,\phi)$. In the present case the space-time is asymptotically flat and necessarily Minkowskian by construction.

The procedure  follows the script established by Bondi  et al. \cite{16}. Thus, four functions are assumed to be expanded as power series in negative powers of $r$, and prescribed on a hypersurface $u=u_0=$constant, namely
 (see \cite{15} for
details) 
\begin{eqnarray}
\gamma&=&cr^{-1} +
\left[C-\frac{1}{6}c^3 -\frac{3}{2}cd^2\right]r^{-3} + Dr^{-4}+...,\label{1}
\end{eqnarray}
\begin{eqnarray}
\delta&=&dr^{-1} +[H+c^2d/2 -d^3/6]r^{-3} + Kr^{-4}+...,\label{2}
\end{eqnarray}
\\
\begin{eqnarray}
F_{12}&=&er^{-2}+(2E+ec+fd)r^{-3}+...,\label{3}
\end{eqnarray}
\begin{eqnarray}
F_{13}\csc\theta&=&fr^{-2}+(2F+ed-fc)r^{-3}+...,\label{4}
\end{eqnarray}
where all coefficients are  functions of $u$, $\theta$ and $\phi$.

Next, from (\ref{1})--(\ref{4}) and a subset of field equations (main equations),  the following expressions are obtained for metric and electromagnetic variables:

\begin{eqnarray}
\beta &=& - ( c^2+d^2) r^{-2}/4 + ..., \label{be1}
\end{eqnarray}
\begin{eqnarray}
U&=&-\left(c_\theta + 2 c \cot{\theta} + d_\phi
\csc{\theta}\right)r^{-2}\nonumber\\
&&+
\left[2 N + 3 (c c_\theta + d d_\theta) + 4 (c^2 + d^2)
\cot{\theta}\right.\nonumber\\
&&- \left.2(c_\phi d - c d_\phi)\csc{\theta}\right]r^{-3} \nonumber\\
&&+\frac{1}{2}\left\{3C_\phi+2C\cot\theta+H_\phi\csc\theta-6(cN+dQ)\right.\nonumber\\
&&- \left.4(2c^2c_\theta+cdd_\theta+c_\theta d^2)-8c(c^2+d^2)\cot\theta-4(c^2d_\phi\right.\nonumber\\ &&\left.+cc_\phi d + 2d^2d_\phi)\csc\theta -(\epsilon e-\mu f)\right\}r^{-4}+\cdots,\label{U2}
\end{eqnarray}
\begin{eqnarray}
W  &=& - (d_\theta + 2 d\cot{\theta} - c_\phi \csc{\theta} )r^{-2}\nonumber\\
 &&+ [ 2Q + 2(c_\theta d - c d_\theta) + 3(c c_\phi + d
d_\phi)\csc{\theta}] r^{-3}\nonumber\\
&&+\frac{1}{2}\left\{3(H_\theta+2H\cot\theta-C_\phi\csc\theta)-6(cQ-dN)\right.\nonumber\\
&&-\left.4(2d^2d_\theta+cc_\theta d+c^2 d_\theta)-8d(c^2+d^2)\cot\theta+4(c_\phi d^2\right.\nonumber\\
&&\left.+cdd_\phi+2c^2c_\phi)\csc\theta+(\mu e+\epsilon f)r^{-4}\right\}
+\cdots,\label{W}
\end{eqnarray}
\begin{eqnarray}
V&=& r - 2M - [N_\theta + N\cot{\theta} + Q_\phi \csc{\theta}
- \frac{1}{2}(c^2 + d^2)\nonumber\\
&&- (c^2_\theta + d^2_\theta) - 4(c c_\theta +
d d_\theta)\cot{\theta}
- 4(c^2+ d^2)\cot^2{\theta}\nonumber\\
 &&- (c^2_\phi + d^2_\phi)\csc^2{\theta} +
4(c_\phi d - c
d_\phi)\csc{\theta}\cot{\theta}\;\;\;\;\;\;\;\;\;\;\;\,\nonumber \\
&&+ 2 (c_\phi
d_\theta - c_\theta d_\phi) \csc{\theta} -\frac{1}{2}(\epsilon^2+\mu^2)] r^{-1}
\nonumber\\
&&+\{\cdots -\mu(f_\theta+f\cot\theta-e_\phi\csc\theta)\}r^{-2}
\cdots,
\label{V2}
\end{eqnarray}
\begin{eqnarray}
F_{01}&=&-\epsilon r^{-2}+(e_\theta + e \cot\theta+f_\phi\csc\theta)r^{-3}+...,
\end{eqnarray}
\begin{eqnarray}
F_{02}&=&X+(\epsilon_\theta-e_u)r^{-1}-\{[E+\frac{1}{2}(ec+fd)]_u\nonumber\\
&&+\frac{1}{2}(e_\theta+e\cot\theta+f_\phi\csc	\theta)]_\theta\}r^{-2}+...,
\end{eqnarray}
\begin{eqnarray}
\csc\theta F_{03}&=&Y+(c_\phi\csc\theta-f_u)r^{-1}-\{[F+\frac{1}{2}(ed-fc)]_u\nonumber\\
&&+\frac{1}{2}(e_\theta+e\cot\theta+f_\phi\csc\theta)]_\phi\csc\theta\}r^{-2}\nonumber\\
&&+\cdots,
\end{eqnarray}
\begin{eqnarray}
\csc\theta F_{23}&=&-\mu-(f_\theta+f\cot\theta-e_\phi\csc\theta)r^{-1}\nonumber\\
&&+\cdots,
\end{eqnarray}
where again, all coefficients  are  functions of $u$, $\theta$ and $\phi$, and subcript letters denote  derivatives.

At this point we have nine functions of three variables which are undetermined  on the initial hypersurface, namely $M, N, Q, \epsilon, \mu, c_u, d_u, X, Y$. However using the remaining field equations (supplementary conditions), equations for the $u$- derivatives of  $M, N, Q, \epsilon, \mu$, in terms of the prescribed functions, can be obtained. Hence if $\gamma, \delta, F_{12}, F_{13}, M, N, Q, \epsilon, \mu$ are prescribed on a given initial hypersurface $u=$ constant, the evolution of the system is fully determined provided the four functions, referrred to as news functions, $c_u, d_u, X, Y$ are given for all $u$. In other words, whatever happens at the source leading to changes in the field, it can only do so by affecting news functions and viceversa. In light of this comment the relationship between the news functions and the occurrence of radition becomes clear.

Furthermore, the mass fuction $m(u)$ which coincides with the Schwarzchild mass in the static case, is defined by
\begin{equation}
m(u)=\int^{2\pi}_0\int^\pi_0 M^{\ast}\sin\theta d\theta d\phi,
\label{m1}
\end{equation}
with 
\begin{eqnarray}
M^{\ast}&=&M+\frac{1}{2} i (\partial/\partial \theta + \cot \theta)(d_\theta +2d \cot \theta-c_\phi \csc \theta)\nonumber \\
&-&\csc \theta \partial/\partial \phi(c_\theta+2c \cot \theta+d_\phi \csc \theta).
\label{m3}
\end{eqnarray}
Then, introducing the intermediate complex quantities:
\begin{equation}
c^{\ast}=c+id, \qquad X^{\ast}=X+i Y,
\label{m4}
\end{equation}
it follows from one of the supplementrary conditions (see \cite{15} for details)
\begin{equation}
m_u=-\int^{2\pi}_0\int^\pi_0( c^{\ast}_u \bar c^{\ast}_u+\frac{1}{2}X^{\ast} \bar X^{\ast})\sin\theta d\theta d\phi,
\label{mas5}
\end{equation}
clearly exhibiting the decreasing of the mass function in the presence of news (bar denotes complex conjugate).

Finally, the following equations derived from the supplementary conditions (Eqs.(14)--(20) in \cite{15}) will be used:
\begin{equation}
M^{\ast}_u=-c^{\ast}_u \bar c^{\ast}_u-\frac{1}{2}X^{\ast} \bar X^{\ast}+\frac{1}{2}\bar {\cal L}_{-1}\bar {\cal L}_{-2} c^{\ast}_u,
\label{vb1}
\end{equation}

\begin{equation}
N^{\ast}_u= {\cal L}_{0} M^{\ast}+2c^{\ast} {\cal L}_{-2} c^{\ast}_u+ {\cal L}_{-2} (c^{\ast}_u c^{\ast})-\bar\epsilon^{\ast} X^{\ast},
\label{vb2}
\end{equation}
\begin{equation}
\epsilon^{\ast}_u= \bar {\cal L}_{-1} \ X^{\ast},
\label{vb3}
\end{equation}

\begin{equation}
4C^{\ast}_u=2c^{\ast 2} \bar c^{\ast}_u+2c^{\ast} \bar M^{\ast}+{\cal L}_{1} N^{\ast}+e^{\ast} X^{\ast},
\label{vb4}
\end{equation}

\begin{eqnarray}
4D^{\ast}_u&=&-\bar {\cal L}_{-3} ({\cal L}_{2} C^{\ast}+2c^{\ast} N^{\ast})\nonumber \\&+&\frac{1}{3}\epsilon^{\ast}{\cal L}_{1} e^{\ast}-\frac{2}{3} c^{\ast}\epsilon{\ast}\bar\epsilon^{\ast}+\frac{4}{3}E^{\ast} X^{\ast},
\label{vb5}
\end{eqnarray}
\begin{equation}
2e^{\ast}_u= - {\cal L}_{0} \bar c^{\ast}-2c^{\ast} \bar X^{\ast},
\label{vb6}
\end{equation}
\begin{equation}
4E^{\ast}_u=-\bar {\cal L}_{-2} ({\cal L}_{1} e^{\ast}-2c^{\ast}\bar\epsilon^{\ast}),
\label{vb7}
\end{equation}
where
\begin{equation}
N^{\ast}=N+i Q, \qquad C^{\ast}=C+i H \qquad D^{\ast}=D+iK,...
\label{vb8}
\end{equation}
\begin{equation}
\epsilon^{\ast}=e+i \mu, \qquad e^{\ast}=e+i f \qquad E^{\ast}=E+iF,...
\label{vb9}
\end{equation}
and
\begin{equation}
{\cal L}_p=-(\partial/\partial \theta-p \cot \theta+i\csc \theta \partial/\partial \phi).
\label{vb10}
\end{equation}

We shall next calculate the general expressions for the vorticity, Poynting (electromagnetic) and super--Poynting vectors for our system.

\section{Vorticity, Poynting and super--Poynting vectors}
The vorticity vector is defined as usual by
\begin{equation}
\omega^\alpha=\frac{1}{2\sqrt{-g}}\eta^{\alpha\eta\iota\lambda}u_\eta
u_{\iota,\lambda},\,\,\eta_{\alpha \beta \gamma \delta} \equiv
\sqrt{-g} \;\;\epsilon_{\alpha \beta \gamma \delta},\label{vv}
\end{equation}
where $\eta_{\alpha\beta\gamma\delta}=+1$ for $\alpha, \beta, \gamma,
\delta$ in even order, $-1$ for $\alpha, \beta, \gamma, \delta$ in
odd order and $0$ otherwise; $u_{\mu}$ is the 4--velocity vector for an observer at rest in the considered frame. The absolute value of $\omega^{\alpha}$ is denoted by $\Omega$, i.e.
\begin{equation}
\Omega=|\omega^{\alpha}\omega_{\alpha}|^{1/2}.
\end{equation}

The electromagnetic Poynting vector is by definition
\begin{equation}
S^\alpha=T^{\alpha\beta}u_\beta,
\end{equation}
whereas  the super-Poynting vector based on the Bel--Robinson tensor, as
defined in \cite{12},
is
\begin{equation}
P_{\alpha}=\eta_{\alpha \beta \gamma
\delta}E^{\beta}_{\rho}H^{\gamma
\rho}u^{\delta},
\label{p1}
\end{equation}
where
$E_{\mu\nu}$ and $H_{\mu\nu}$, are the electric and magnetic parts of
Weyl
tensor, respectively, formed from Weyl tensor $C_{\alpha
\beta \gamma
\delta}$ and its dual
$\tilde C_{\alpha \beta \gamma \delta}$ by
contraction with the four
velocity vector, i.e.
\begin{equation}
E_{\alpha \beta}=C_{\alpha \gamma \beta
\delta}u^{\gamma}u^{\delta},
\label{electric}
\end{equation}
\begin{equation}
H_{\alpha \beta}=\tilde C_{\alpha \gamma \beta
\delta}u^{\gamma}u^{\delta}=
\frac{1}{2}\eta_{\alpha \gamma \epsilon
\delta} C^{\epsilon
\delta}_{\quad \beta \rho}
u^{\gamma}
u^{\rho}.
\label{magnetic}
\end{equation}

Let us
first calculate the vorticity for the congruence
of observers at rest in
(\ref{bsm}), whose four--velocity vector is given
by  $u^{\alpha} =A^{-1}\delta^{\alpha}_{u}$,
where  $A$ is given
by
\begin{eqnarray}
A&=&(Vr^{-1}e^{2\beta}-r^2 e^{2\gamma}U^2\cosh 2\delta\nonumber\\
&-&r^2 e^{-2\gamma} W^2
\cosh 2\delta -2r^2 UW\sinh
2\delta)^{1/2}.
\end{eqnarray}
Thus, (\ref{vv}) lead us to
\begin{equation}
\omega^\alpha=(\omega^u,\omega^r,\omega^\theta,\omega^\phi),
\end{equation}
where
\begin{widetext}
\begin{eqnarray}
\omega^u&=&-\frac{1}{2A^2\sin\theta}
\{r^2e^{-2\beta}(WU_{r}-UW_{r})
+\left[2r^2\sinh 2\delta
\cosh 2\delta (U^2e^{2\gamma}+W^2e^{-2\gamma}) 
+ 4UWr^2\cosh^2 2\delta\right]e^{-2\beta}\gamma_{r}
\nonumber \\&+&2r^2e^{-2\beta}(W^2e^{-2\gamma}
-U^2e^{2\gamma})\delta_{r}
+e^{2\beta}[e^{-2\beta}(U\sinh
2\delta+e^{-2\gamma} W\cosh
2\delta)]_{\theta}\nonumber \\
&-&e^{2\beta}[e^{-2\beta}(W\sinh
2\delta+e^{-2\gamma} U\cosh
2\delta]_{\phi}\},
\end{eqnarray}
\end{widetext}
\begin{widetext}
\begin{eqnarray}
\omega^r&=&\frac{1}{e^{2\beta}\sin\theta}\{2r^2 A^{-2}[ (
(U^2e^{2\gamma}+W^2e^{-2\gamma})\sinh
2\delta \cosh 2\delta
+UW\cosh^2 2\delta) \gamma_{u})+(W^2e^{-2\gamma}-U^2e^{2\gamma})\delta_{u}
\nonumber \\&+&\frac{1}{2}(WU_{u}-UW_{u})]
+A^2[A^{-2}(We^{-2\gamma}\cosh 2\delta+U\sinh 2\delta)]_{\theta} 
-A^2[A^{-2}(W\sinh 2\delta+Ue^{2\gamma}\cosh 2\delta)]_{\phi}\},
\end{eqnarray}
\end{widetext}
\begin{widetext}
\begin{eqnarray}
\omega^\theta&=&\frac{1}{2
r^2\sin\theta}\{A^2e^{-2\beta}[r^2A^{-2}(U\sinh
2\delta+We^{-2\gamma}\cosh
2\delta)]_{r}
 -e^{2\beta} A^{-2}[e^{-2\beta} r^2(U\sinh
2\delta+e^{-2\gamma} W\cosh
2\delta)]_{u}\nonumber \\
&+&e^{2\beta} A^{-2}(e^{-2\beta} A^2)_{\phi} \},
\end{eqnarray}
\end{widetext}
and
\begin{widetext}
\begin{eqnarray}
\omega^\phi&=&\frac{1}{2r^2\sin\theta} \{
A^2e^{-2\beta}[r^2 A^{-2} (W\sinh 2\delta+Ue^{2\gamma}\cosh
2\delta)]_{r} 
-e^{2\beta} A^{-2}[r^2e^{-2\beta}(W\sinh
2\delta+Ue^{2\gamma}\cosh
2\delta)]_{u}\nonumber \\
&+&A^{-2}e^{2\beta}(A^2e^{-2\beta})_{\theta}\}.
\end{eqnarray}
\end{widetext}
Although algebraic manipulation by hand is feasible for the Bondi
metric, for the Bondi--Sachs one is quite cumbersome. {Therefore we write a Maple 15 script (available upon request) which uses 
intrinsic procedures to deal with tensors. We proceed in two steps. First, we calculate and save (\ref{electric}), (\ref{magnetic}) and (\ref{p1}), using the general form (\ref{metric}), that is, without using the metric function expansions (\ref{be1})--(\ref{V2}). Second, we expand all the relevant objects separately (metric, Weyl, electric and magnetic parts) up to the leading order. After these two simple steps we were able to write the output for the super--Poynting.}

We use the shift vector $U^A=(U,W/\sin\theta)$
and the 2--surfaces metric of constant $u$
\begin{equation}
h_{AB}=\left (
\begin{array}{ll}

e^{2\gamma}\cosh{2\delta}&\,\,\,\sinh{2\delta}\sin\theta \\

\sinh{2\delta}\,\sin\theta &\,\,\,
e^{-2\gamma}\cosh{2\delta}\,\sin^2\theta
               \end{array}
\right ).
\end{equation}
In our calculations we keep these auxiliary variables, $U^A$ and $h_{AB}$, as far as was possible.

Thus, for the absolute value of $\omega^\mu$ we
get
\begin{eqnarray}
\Omega&=&\Omega_\mathcal{G}r^{-1}+\cdots+\Omega_{\mathcal{GEM}}r^{-3}+\cdots,
\end{eqnarray}
where subscripts $\mathcal{G}$, $\mathcal{GEM}$ and $\mathcal{EM}$ stand for gravitational, gravito--electromagnetic  and electromagnetic. At this point a remark on the meaning of this notation is in order:  ``gravitational'' terms refer to those terms containing exclusively functions $M, N, Q, c_u, d_u$ and their derivatives. ``Electromagnetic'' terms are those containing exclusively  functions $\epsilon, \mu, X, Y$ and their derivatives, whereas ``gravitomagnetic'' terms refer to those containing functions of either kind and/or combination of both. It should be clear that all functions are related through field equations, and therefore the established division is rather formal, however the splitting as indicated is useful for the discussion below.

In order to get some insight  let us first consider the axially and reflection symmetric case, i.e., $\partial/\partial \phi =d=f=H=K=F=Y=Q=0$. Thus, we obtain 
{
\begin{eqnarray}
\Omega_{\mathcal{GEM}}&=&-\frac{1}{4}[4c_uMc_\theta+2N_{\theta\theta}+2c_uN+3C_{\theta u}+2N\nonumber\\
&&-4(Mc)_\theta-(\epsilon e)_u+2(4M-c)N_u\nonumber\\
&&-4cc_\theta c_u+8M_\theta M-2(1+2c_u)N\cot^2\theta\nonumber\\
&&+(8cM-8M^2-2N_\theta-3c^2) c_{\theta u}\nonumber\\
&&+2(3c_uc^2-2c_uN_\theta-8c_uM^2+4c_ucM \nonumber\\
&&-Nc_{\theta u}-4cM+3c_u+N_\theta)\cot\theta],
\end{eqnarray}
}
where we observe the contribution $\frac{1}{4}(\epsilon e)_u$, which is purely electromagnetic.  This latter term together with terms proportional to $N_u$ and $C_u$ may in turn be expressed through electromagnetic news ($X^{\ast}$) by means of  (\ref{vb2})--(\ref{vb4}), and (\ref{vb6}). Doing so we clearly identify electromagnetic radiation as a vorticity source.

For the angular contravariant super--Poynting components (in the axially symmetric case) we identify many $\mathcal{GEM}$ contributions. One of them is explicit, e.g.
 $c_{uu}(e\epsilon)_u$ in $P^{\theta}$ (the complete term is too long to display here; see the appendix) whereas other terms appear by replacing $u$ derivatives of metric variables by means of (\ref{vb1})--(\ref{vb10}). It is remarkable the following simplification
\begin{equation}
P^{\phi}_\mathcal{GEM}=-\csc\theta c_{uu}(\mu e)_u.
\end{equation}

Let us now get back to the most general three--dimensional case, where {except} for the leading term,
\begin{eqnarray}
\Omega_\mathcal{G}&=&-\frac{1}{2}\left[(c_{\theta u}+2c_{u}\cot\theta +
d_{\phi u} \csc\theta)^2\right.\nonumber\\
&+&\left.(d_{\theta u}+2d_{u}\cot\theta - c_{\phi u}
\csc\theta)^2\right]^{1/2},\label{On}
\end{eqnarray}
the next leading gravito--electromagnetic term is too long to write here or anywhere (the expression is available upon request). 

However, the important point to stress here is that  $\Omega_{\mathcal{GEM}}$ {contains} electromagnetic news functions ($X^{\ast}$) by means of, for instance  of terms 
\begin{equation}
3\sin\theta c_{\theta u}^2d_{\phi u}e\epsilon_u,\qquad 3c_{\theta u}d_{\phi u}^2f\mu_u,\label{d1}
\end{equation}
which are typical and selected contributing terms to $\Omega_{\mathcal{GEM}}$.

Next, we write the electromagnetic Poynting vector as
\begin{equation}
S^u=S^u_\mathcal{EM}r^{-4}+S^u_\mathcal{GEM}r^{-5}\cdots,
\end{equation}
\begin{equation}
S^r=S^r_\mathcal{EM}r^{-2} + S^r_\mathcal{GEM} r^{-3}+\cdots,
\end{equation}
\begin{equation}
S^\theta=S^\theta_\mathcal{EM} r^{-4}+S^\theta_\mathcal{GEM} r^{-5}+\cdots,
\end{equation}
\begin{equation}
S^\phi=S^\phi_\mathcal{EM}r^{-4}+S^\phi_\mathcal{GEM} r^{-5}+\cdots,
\end{equation}
where
\begin{equation}
S^u_\mathcal{EM}=\frac{1}{2}(\mu^2+\epsilon^2),
\end{equation}
{\begin{eqnarray}
S^u_\mathcal{GEM}&=&(\mu f -\epsilon e )\cot\theta 
-(\epsilon f_\phi+\mu e_\phi)\csc\theta\nonumber
 \nonumber \\&&+\mu f_\theta+ MS^u_\mathcal{EM}-\epsilon e_\theta,
\end{eqnarray}}
\begin{equation}
S^r_\mathcal{EM}=X^2+Y^2,\\
\end{equation}
{\begin{eqnarray}
S^r_\mathcal{GEM}&=&-2 Xe_u+ MS^r_\mathcal{EM}+2 X\epsilon_\theta
-4 dXY \nonumber \\&&
+2 c(Y^2-X^2)-2 Yf_u+2Yc_\phi\csc\theta,\\
S^\theta_\mathcal{EM}&=&\epsilon X+\mu Y,\\
S^\theta_\mathcal{GEM}&=&-(Ye_\phi+Xf_\phi-\mu c_\phi)\csc\theta-(Xe-Yf)\cot\theta
\nonumber\\&&+MS^\theta_\mathcal{EM}-2\epsilon (Yd+Xc)+\epsilon\epsilon_\theta- Xe_\theta-\mu f_u\nonumber\\
&&+ Yf_\theta-\epsilon e_u,
\end{eqnarray}}
\begin{equation}
S^\phi_\mathcal{EM}=\csc\theta(\epsilon Y-\mu X),
\end{equation}

{\begin{eqnarray}
S^\phi_\mathcal{GEM}&=&-\csc\theta[(Ye+Xf)\cot\theta-MS^\phi_\mathcal{EM}\sin\theta\nonumber\\&&-(Xe_\phi-Yf_\phi+\epsilon c_\phi)\csc\theta+(2d\epsilon
+f_\theta)X\nonumber\\
&&+(e_\theta-2c \epsilon)Y+\epsilon f_u+(\epsilon_\theta-e_u)\mu].
\end{eqnarray}}
It is worth noticing that terms explicitly containing $X^{\ast}$ appear in $\theta$ and $\phi$ component of  the Poynting vector.

Next, calculation of the super--Poynting gives the
following
result
\begin{equation}
P_{\mu}=(0, P_r, P_{\theta}, P_{\phi}),
\end{equation}
where
the explicit terms are too long to be written at this point.

The leading terms for each super--Poynting (contravariant) component are
\begin{eqnarray}
P^u&=&P^{u}_{\mathcal{G}}r^{-4}+\cdots, \nonumber\\ \label{sp0}
P^r&=&P^r_{\mathcal{G}}r^{-4}+\cdots, \nonumber\\ \label{sp1}
P^\theta &=& P^{\theta}_{\mathcal{G}}r^{-4} +\cdots+ P^{\theta}_{\mathcal{GEM}}r^{-6}+\cdots, \nonumber\\\label{sp2}
P^\phi &=& P^{\phi}_{\mathcal{G}}r^{-4} +\cdots+ P^{\phi}_{\mathcal{GEM}}r^{-6}+\cdots, \label{sp3}
\end{eqnarray}
where
{
\begin{eqnarray}
P^{\theta}_{\mathcal{G}}&=&2[2(d_{uu}d_u+c_{uu}c_u)\cot\theta+c_{uu}c_{\theta u}\nonumber\\
&&+(c_{uu}d_{\phi u}-d_{uu}c_{\phi u})\csc\theta+d_{u u}d_{\theta u}],\\
P^{\phi}_{\mathcal{G}}&=&2\csc\theta[2(c_{uu}d_u-d_{u u}c_u)\cot\theta+c_{uu} d_{\theta u}\nonumber\\
&&-d_{uu}c_{\theta u}-(c_{uu}c_{\phi u}+d_{u u}d_{\phi u})\csc\theta].
\end{eqnarray}}
Other terms are too long to display. However the important point is that, again, terms containing electromagnetic news appear in $P^{\mu}_{\mathcal{GEM}}$, as for
example
\begin{equation}
 \cot^4\theta c_{uu}e\epsilon_u, \qquad
 \cot^4\theta c_{uu}f\mu_u \label{d2}
\end{equation}
and
\begin{equation}
 2\csc^3\theta \cot^2\theta d_{u u}e\epsilon_u,\quad \csc^5\theta d_{u u}f\mu_u,\label{d3}
\end{equation}
which are typical and selected contributing terms to the super--Poynting components $P^{\theta}_{\mathcal{GEM}}$ and $P^{\phi}_{\mathcal{GEM}}$ respectively.

It is easy to check that the stationary case satisfy well known results,
for instance, that of Bonnor or the Kerr--Newman. 
Now, contributions to each relevant object are superior for $_{s}\Omega$ and $_{s}P^\alpha$ with respect to the general (radiative) case, but it is not true for $_{s}S^\alpha$ which keeps the same leading terms:
\begin{eqnarray}
_{s}\Omega&=&_{s}\Omega_\mathcal{G}r^{-2}+\cdots + {_{s}\Omega}_{\mathcal{GEM}}r^{-4}+\cdots \\
_{s}S^u&=&_{s}S^{u}_{\mathcal{EM}}r^{-4}+{_{s}S}^{u}_{\mathcal{GEM}}r^{-5}+\cdots \\\
_{s}S^r&=&_{s}S^r_{\mathcal{GEM}}r^{-4}+\cdots \\
_{s}S^\theta&=&_{s}S^\theta_{\mathcal{GEM}}r^{-5}+\cdots \\
_{s}S^\phi&=&_{s}S^\phi_{\mathcal{GEM}}r^{-5}+\cdots\\
_{s}P^u&=&_{s}P^{u}_\mathcal{G}r^{-6}+\cdots, \label{sp0e}\\
_{s}P^r&=&_{s}P^{r}_\mathcal{G}r^{-6}+\cdots, \label{sp1e} \\
_{s}P^\theta &=& _{s}P^{\theta}_\mathcal{G}r^{-7} +\cdots+ {_{s}P}^{\theta}_{\mathcal{GEM}}r^{-9}+\cdots, \label{sp2e}
\\
_{s}P^\phi &=& _{s}P^{\phi}_\mathcal{G}r^{-7} +\cdots+ {_{s}P}^{\phi}_{\mathcal{GEM}}r^{-9}+\cdots. \label{sp3e}
\end{eqnarray}
where
\begin{widetext}
{
\begin{eqnarray}
_{s}\Omega_\mathcal{G}&=&  \frac{1}{2}\{[4(d_{\phi\phi}c_{\phi\theta}-3d_\theta c_\phi)\cos\theta+2(4c_{\phi\theta} c_\phi-3d_\theta d_{\phi\phi})]\cos\theta+[4(dd_{\phi\phi}-dd_{\theta\theta}+M^2_\phi+M^2_\theta+c^2_{\phi\theta}+d^2)\nonumber\\
&&+(d_{\phi\phi}-d_{\theta\theta})^2]\csc\theta+[2(2c_{\phi\theta}-3d_\theta\cos\theta)(2d-d_{\theta\theta})+4(2dc_\phi-3c_{\phi\theta}d_\theta-d_{\theta\theta}c_\phi)\cot\theta+(9d^2_\theta-d^2_{\theta\theta}\nonumber\\
&&+4(dd_{\theta\theta}-d^2-M^2_\theta))\cos\theta\cot\theta]\sin^2\theta +[4d_{\phi\phi}c_\phi+(4(dd_{\theta\theta}-d^2-M^2_\theta-dd_{\phi\phi}-M^2_\phi-c^2_{\phi\theta}+c^2_\phi)\nonumber\\
&&-d^2_{\theta\theta}+2d_{\theta\theta}d_{\phi\phi})\cos\theta]\cot\theta\}^{1/2}
\end{eqnarray}}
\end{widetext}
(the expression for $_{s}\Omega_{\mathcal{GEM}}$ is too long to write here or anywhere; we check that is manifestly gravito--electromagnetic)
{
\begin{equation}
_{s}S^{u}_{\mathcal{EM}}=S^{u}_{\mathcal{EM}}
\end{equation}
\begin{eqnarray}
_{s}S^{u}_{\mathcal{GEM}}&=&(\mu f -\epsilon e )\cot\theta \nonumber\\
&&-(\epsilon f_\phi+\mu e_\phi)\csc\theta\nonumber\\
&&+\mu f_\theta-\epsilon e_\theta+ MS^{u}_{\mathcal{EM}}
\end{eqnarray}}
\begin{equation}
_{s}S^r_{\mathcal{GEM}}=\epsilon^2_\theta+
 c_\phi^{2}\csc^2\theta
 \end{equation}
 \begin{equation}
_{s}S^\theta_{\mathcal{GEM}}=\epsilon_\theta\epsilon+c_\theta\mu\csc\theta
\end{equation}
\begin{equation}
_{s}S^\phi_{\mathcal{GEM}}=(c_\phi\epsilon\csc\theta-\epsilon_\theta\mu)\csc\theta
\end{equation}
(coefficients for $_{s}P^\alpha$ are too long to display anywhere).

{The above expressions illustrate the Bonnor--like mechanism for any stationary electrovacuum solution (see \cite{10} for further discussion).}
\section{Discussion}
{We have seen that electromagnetic radiation as described by electromagnetic news functions does produce vorticity. {It is important to stress that this is so even in the case of minimum electromagnetic degrees of freedom (which is one), the reflection and axially
symmetric  case, when $X\neq 0$ and $Y=0$, implying that the
above mentioned effect is generic.} Since vorticity unavoidably produces frame dragging, we have established the link between these two physical effects.  Furthermore we have identify the presence of electromagnetic news both in the Poynting and the super--Poynting components orthogonal to the vorticity vector. Doing so we have proved that a Bonnor--like mechanism is at work in this case too. However it is important to emphasize that  in the  present situation vorticity is generated by the contributions of, both,  the Poynting   and  the super--Poynting vectors,   on the planes orthogonal to the vorticity vector.} {It must be stressed that the mechanism to explain how electromagnetic and gravitational radiation produce vorticity invokes the concept of superenergy (its flow), which is one of the most important concepts in general relativity involving the congruence of observers \cite{h11}.}

Before proceeding further with our discussion two comments are in order:
\begin{itemize}
\item We have clearly established the link between vorticity and electromagnetic radiation, which as mentioned before implies a link between electromagnetic radiation and frame dragging. This was the main goal of our work. However we have not calculated in detail the  resulting precession rate of a falling gyroscope under such a  circumstance, since it is out of the scope of this manuscript. It goes without saying that for an explicit experiment setup  such a calculation should be provided.

\item It should be clear that the magnitude of the mentioned effect is directly related to the intensity of the electromagnetic emission  ($Y^\ast$).
\end{itemize}

{Simulations from numerical relativity could shed some light to figure out how to measure the effect reported here. First, in the study of binary black holes dynamics near electromagnetic fields and plasmas \cite{palenzuelaetal}, it was displayed how the system imprints characteristics on the two induced wavebands. In the present case, as an inverse problem, the electromagnetic and gravitational radiation produce precession (on test gyroscopes) which has to be imprinted by the waves. Second, almost in the same aforementioned context, it was possible to track the precession of compact binaries from gravitational wave signals \cite{shha11}, locating the frame from which the ($l=2$, $|m|=2$) modes are maximized. We suppose that in the same way as the ``quadrupolar--aligned" frame is located, the ``dipolar--aligned" frame could be find from electromagnetic modes. This simple method can be applied to the ensuing gyroscope precession, as reported in here.}

{The potential observational consequences of the reported effect should be seriously considered.}  Indeed, as we mentioned before, intense electromagnetic outbursts are expected  from hyperenergetic phenomena such as collapsing hypermassive neutron stars and Gamma Ray Bursts (see \cite{20}, \cite{21} and references therein). Although we are not able at the present to estimate the required sensitivity of  the gyroscope to measure such an effect, the high intensity of radiation in the above mentioned scenarios  leaves open  the question about its detection with present technology. {In  fact, the   direct experimental evidence of the existence of the Lense--Thirring effect \cite{22, ciu1, ciu2} brings out the high degree of development achieved in  the required  technology. In the same direction point recent proposals to detect frame dragging by means of ring lasers \cite{rl1, rl1bis, rl2, rl3, rl4}}

Finally we would like to mention that frame dragging produced by the so called optical vortices has been recently described in the linear regime \cite{23}. However it should be observed that the effect reported here  stems from non--linear terms, as it can be seen from (\ref{d1}), (\ref{d2}) and (\ref{d3}), bringing out the relevance of nonlinearities in the general relativistic description of radiation.
\begin{acknowledgments}
WB thanks to Departamento de F\'\i sica Te\'orica e Historia de la Ciencia, Universidad del Pa\'\i s Vasco, for hospitality, specially to J. 
Ib\'a\~nez and A. Di Prisco; also to Intercambio Cient\'\i fico Program, U.L.A., for financial support. 
\end{acknowledgments} 
\begin{widetext}
\appendix*
\section{Expression for the axially and reflection symmetric $P^\theta_\mathcal{GEM}$} 
{
\begin{eqnarray}
P^{\theta}_\mathcal{GEM}&=&\frac{1}{4}
[3c_\theta c_{\theta\theta u} c_u-4c_{\theta u} c_uc_{\theta\theta}-4 c_{uu}\epsilon_ue-24 c_{uu}cN_u-4 c_{uu}\epsilon e_u+14cc_{\theta u} c_u-60c_{\theta u} c_{uu}M^2-32M c_{\theta u}c_u\nonumber\\
&&-18c_{uu}cc_\theta-12c_{\theta u} c_{uu}N_\theta-40c_{\theta u} c_{uu}c^2-24M_\theta cc_{uu}+8cc_{\theta u}M_u-17cc_{\theta u} c_{\theta\theta u}+12M c_{\theta u}c_{\theta\theta u}-16c_{uu}M c_\theta\nonumber\\
&&+36M_\theta c_{uu}M+30M c_{\theta u}cc_{uu}-18c_{uu}M c_\theta c_u-112c_uc_\theta cc_{uu}-24c_{uu}N c_u+8 c_{uu}C_{\theta u}+12M_\theta c_u
-8c_{\theta u} N_{\theta u}
\nonumber\\
&&-4c_{\theta u}M_{\theta\theta}-16 c_{uu}N-12M c_{\theta u}-6c_u^2c_\theta+8 c_{uu}N_{\theta\theta}-8c_{\theta u} c_{uu}-16 c_\theta c_{\theta u}^2-6M_\theta c_{\theta\theta u}+(16c_ucM_u-24c_{uu}c^2\nonumber\\
&&-24Mc_u-16c_uC_{uu}-48c_{uu}Mc_uc-120c_{uu}M^2c_u-2c_{\theta u}M_\theta+8c_{\theta u}N_u-25cc_{\theta u}^2-12c_{uu}Nc_{\theta u}-8cc_u^2\nonumber\\
&&+12Mc_{\theta u}^2-8c_uM_{\theta\theta}-32c_{uu}cM+24c_{\theta\theta u}Mc_u-16c_uN_{\theta u}-8c_u^2c_{\theta\theta}+8c_{uu}N_\theta-64c_u^2M-41c_{\theta u}c_\theta c_u\nonumber\\
&&+16c_{uu}C_u-160c_{uu}c_uc^2-16c_ucc_{\theta\theta u}-24c_{uu}c_uN_\theta)\cot\theta].\nonumber
\end{eqnarray}}
\end{widetext}

\end{document}